\documentclass[final]{article}

\usepackage{graphicx}

\def\aa{{A\&A}}

\def\aj{{AJ}}
\def\annrev{{ARA\&A}}
\def\apj{{ApJ}}
\def\apjs{{ApJS}}

\def\mnras{{MNRAS}}

\def\prd{{Phys. Rev. D}}
\def\lesssim{\mathrel{\hbox{\rlap{\hbox{\lower4pt\hbox{$\sim$}}}\hbox{$<$}}}}
\def\gtrsim{\mathrel{\hbox{\rlap{\hbox{\lower4pt\hbox{$\sim$}}}\hbox{$>$}}}}

\begin{document}

\title{\bf Relativistic Collapse of Rotating Supermassive Stars to Supermassive 
Black Holes}

\author{{\large Stuart L. Shapiro}\\
{\it University of Illinois at Urbana-Champaign}}
\date{}
\pagestyle{empty}
\maketitle

\begin{abstract}
There is compelling evidence that supermassive black holes (SMBHs) 
exist. Yet the origin of these objects, or their seeds, is still unknown.
We are performing general relativistic simulations of gravitational 
collapse to black holes in different scenarios to help reveal how 
SMBH seeds might arise in the universe. SMBHs with $ \sim 10^9 ~M_{\odot}$ 
must have formed by $z > 6$, or within $10^9$ yrs after 
the Big Bang, to power quasars. It may be difficult for 
gas accretion to build up such a 
SMBH by this time unless the initial seed black hole already
has a substantial mass. One plausible progenitor of a massive seed 
black hole is a supermassive star (SMS). We have followed the collapse 
of a SMS to a SMBH by means of 3D hydrodynamic simulations in 
post-Newtonian gravity and axisymmetric simulations 
in full general relativity.  The initial SMS of arbitrary mass M in these 
simulations rotates uniformly at the mass--shedding limit and 
is marginally unstable to radial collapse. The final black hole mass and 
spin are determined to be $M_{\rm h}/M \approx 0.9$ and 
$J_{\rm h}/M_{\rm h}^2 \approx 0.75$. 
The remaining 
mass goes into a disk of mass $M_{\rm disk}/M \approx 0.1$. 
This disk arises even 
though the total spin of the progenitor star, $J/M^2 = 0.97$, is 
safely below the Kerr limit.  The collapse generates a mild 
burst of gravitational radiation. Nonaxisymmetric bars or one-armed spirals 
may arise during the quasi-stationary evolution of a SMS, during its collapse, 
or in the ambient disk about the hole, and are 
potential sources of quasi-periodic waves, detectable by LISA.  
\end{abstract}

\thispagestyle{empty}

\section*{\bf Introduction}

There is substantial evidence that supermassive black holes (SMBHs) of mass 
$\sim 10^6 - 10^{10} ~M_{\odot}$ exist and are the engines that power active 
galactic nuclei (AGNs) and quasars \cite{rees84,rees98,rees01,mach99}.  There 
is also ample evidence that SMBHs reside at the centers of many, and perhaps 
most, galaxies \cite{rich98,ho99}, including the Milky Way 
\cite{genz97,ghez00,scho02}. 

\begin{figure}[tb]
\centering
\includegraphics[height=7cm]{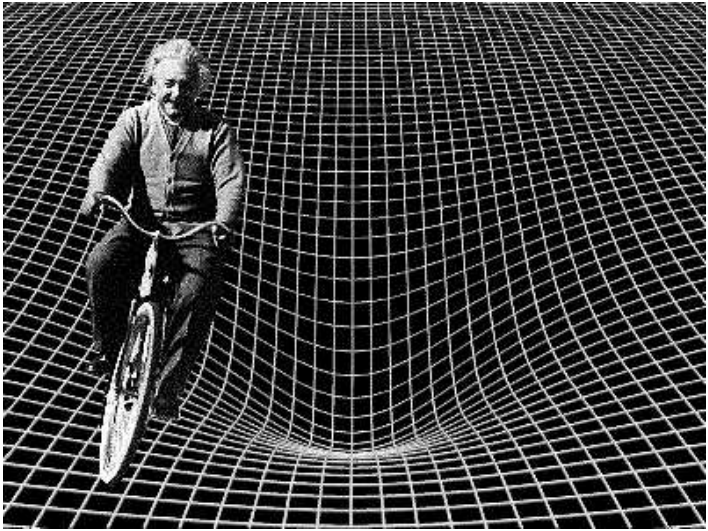}
\caption{The formation of a black hole is a strong-field gravitational
phenomenon in curved spacetime that requires Einstein's equations of general
relativity for a description and, in nontrivial cases, numerical simulations
for a solution.}
\label{eins}
\end{figure}

Since quasars have been discovered out to redshift $z \gtrsim 6$ 
\cite{fan00, fan01}, the first SMBHs must have formed by $z_{\rm BH} \gtrsim 6$, or 
within $t_{\rm BH} \lesssim 10^9$ yrs after the Big Bang.  However, the 
cosmological origin of SMBHs is not known. This issue remains one of the 
crucial, unresolved aspects of structure formation in the early universe.
Gravitationally, black holes are strong-field objects whose properties are 
governed by Einstein's theory of relativistic gravitation --- general 
relativity.  General relativistic simulations of gravitational collapse to 
black holes therefore may help reveal how, when and where SMBHs, or their 
seeds, form in the universe. Simulating plausible paths by which the 
very first seed black holes may have arisen comprises a timely 
computational challenge (see Fig.~\ref{eins}). It is an area
in which the tools of numerical relativity may be exploited to 
address a fundamental question in cosmology. Performing such simulations
may also help identify plausible astrophysical scenarios and sites for
promising gravitational wave sources involving black holes.


We are actively developing new algorithms and new computer 
codes to solve Einstein's
field equations of general relativity, coupled to the equations of relativistic 
hydrodynamics, in three spatial dimensions plus time (3+1). As our codes 
have come online, we have applied them to explore
different astrophysical scenarios involving strong gravitational
fields and the generation of gravitational waves. These have included 
the inspiral and coalescence of binary neutron stars and binary 
black holes, the growth of
instabilities in rotating stars, the nonlinear evolution of unstable
stars, and the collapse and collision of rotating stars and rotating clusters of
collisionless matter, to name
a few. [For recent reviews of simulations 
of compact binary stars and references, see \cite{rasio99} and 
\cite{baum03}]. We have also  
performed a wide range of simulations in recent years to study
alternative scenarios leading to 
the formation of supermassive black holes, or their seeds. 
[For a review of these calculations 
and references, see \cite{shap03}] 
Here we will focus on one specific 
scenario, namely, the collapse of a rotating, supermassive star (SMS) to
a supermassive black hole.

\section*{The Onset of Dynamical Instability}

The formation of SMBHs through the growth of black hole 
seeds by gas accretion is supported by the
consistency between the total energy density in QSO light and the SMBH mass
density in local galaxies, adopting a reasonable accretion
rest-mass--to--energy conversion efficiency \cite{sol82, yu02}.
But SMBHs must be present by $z_{\rm BH} \gtrsim 6$ to power quasars.  
It has been argued \cite{gned01} that if they grew by accretion 
from smaller 
seeds, substantial seeds of mass $\gtrsim 10^5 ~M_{\odot}$ must already 
be present at $z \approx 9$ to have had sufficient time to build 
up to a typical quasar black hole mass of 
$\sim 10^9 ~M_{\odot}$. A likely progenitor is a very massive object (e.g., an 
SMS) supported by radiation pressure. 


SMSs ($10^3 \lesssim M/M_{\odot} \lesssim 10^{13}$) may form when contracting 
or colliding primordial gas builds up sufficient radiation pressure to inhibit 
fragmentation and prevent star formation (see, e.g., \cite{brom03}).  
SMSs supported by radiation pressure will evolve in a quasi-stationary
manner to the point of onset of dynamical collapse due to general relativity
\cite{chandra64a, chandra64b, feyn}
Unstable SMSs with $M \gtrsim 10^5 ~M_{\odot}$ and metallicity 
$Z \lesssim 0.005$ do not disrupt due to thermonuclear explosions during 
collapse \cite{full86}. In fact, recent Newtonian 
simulations suggest that evolved zero-metallicity (Pop III) stars  
$\gtrsim 300 ~M_{\odot}$ do not disrupt but collapse with negligible mass loss 
\cite{fry01}. This finding could be important since the 
first generation of stars may form in the range $10^2 - 10^3 ~M_{\odot}$ 
\cite{brom99, abel00}. A combination of 
turbulent viscosity and magnetic fields likely will keep a spinning SMS in 
uniform rotation (\cite{bisno67, wag69, zel71, shap00};
but see \cite{new01} and the final section below for an 
alternative). As they cool and contract, uniformly rotating SMSs reach the 
maximally rotating {\it mass-shedding limit}\ and subsequently evolve in a 
quasi-stationary manner along a mass-shedding sequence until reaching the 
instability point.  At mass-shedding, the matter at the equator moves in a 
circular geodesic with a velocity equal to the local Kepler velocity 
\cite{baum99}. 

It is straightforward to understand the radial instability induced by general
relativity in a SMS by using an energy variational principle \cite{zel71, shap83}.
Let $E=E(\rho_c)$ be the total 
energy of a momentarily static, spherical fluid configuration characterized by 
central mass density $\rho_c$.  The condition that $E(\rho_c)$ be an extremum 
for variations that keep the total rest mass and specific entropy distribution 
fixed is equivalent to the condition of hydrostatic equilibrium and 
establishes the relation between the equilibrium mass and central density:
\begin{equation} \label{1var}
\frac {\partial E}{\partial \rho_c} = 0 \ \Longrightarrow \ M_{\rm eq} 
= M_{\rm eq}(\rho_c) \\ ~~~~~({\rm equilibrium}).
\end{equation}
The condition that the second variation of $E(\rho_c)$ be zero is the 
criterion for the onset of dynamical instability.  This criterion shows that 
the turning point on a curve of equilibrium mass vs. central density marks the 
transition from stability to instability:
\begin{equation} \label{2var}
\frac{{\partial}^2 E}{\partial {\rho_c}^2} = 0 \ \Longleftrightarrow \
\frac{\partial M_{\rm eq}}{\partial \rho_c} = 0 \\ 
~~~~~({\rm onset\ of\ instability}).
\end{equation}

\begin{figure}[tb]
\hspace*{2cm}\includegraphics[angle=-90,width=11cm]{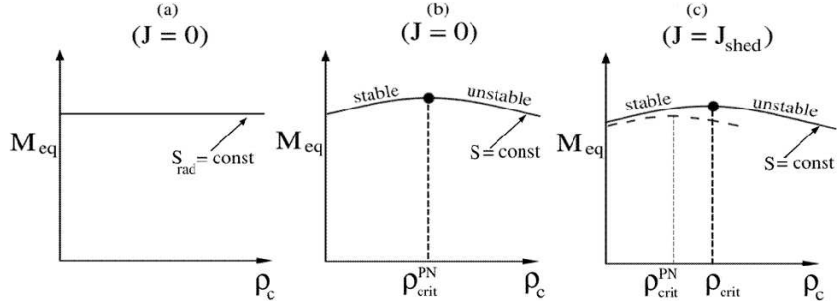}
\caption{A sketch of mass versus central density along an equilibrium
sequence of SMSs of fixed entropy. Panel ({\it a}) shows nonrotating, 
spherical Newtonian models supported by pure radiation pressure; ({\it b}) 
shows nonrotating, spherical PN models supported by radiation pressure plus 
thermal gas pressure; ({\it c}) shows rotating PPN models spinning at the 
mass-shedding limit.}
\label{sketch}
\end{figure}

Consider the simplest case of a spherical Newtonian SMS supported solely by 
radiation pressure and endowed with zero rotation.  This is an $n=3$, 
($\Gamma = 1+1/n= 4/3$) polytrope, with pressure 
\begin{equation} \label{eos1}
P = P_{\rm rad} = \frac{1}{3} a T^4 = K \rho^{\frac{4}{3}},
\end{equation}
where $K = K(s_{\rm rad})$ is a constant determined by the value of the 
(constant) specific entropy $s_{\rm rad} = \frac{4}{3} a T^3/n$  in the star. 
Here $T$ is the temperature, $n$ is the baryon number density, and $a$ is the 
radiation constant.  Consider a sequence of configurations with the same 
specific entropy but different values of central density. The total energy of 
each configuration is 
\begin{equation} \label{en1}
E(\rho_c) = U_{\rm rad} + W,
\end{equation}
where $U_{\rm rad}$ is the total internal radiation energy and $W$ is the 
gravitational potential energy.  Applying the equilibrium 
condition~(\ref{1var}) to this functional yields
$M_{\rm eq} = M_{\rm eq}(s_{\rm rad})$, i.e. the equilibrium mass depends only 
on the specific entropy and is independent of central density 
(see Fig.~\ref{sketch}{\it a}).  Applying the stability condition~(\ref{2var}) 
then shows that all equilibrium models along this sequence are marginally 
stable to collapse.


Now let us account for the effects of general relativity.  If we include the 
small (de-stabilizing) Post-Newtonian (PN) correction to the gravitational 
field, we must also include a comparable (stabilizing) correction to the 
equation of state arising from thermal gas pressure:
\begin{equation} \label{eos2}
P = P_{\rm rad} + P_{\rm gas} = \frac{1}{3} a T^4 + 2nkT,
\end{equation}
where we have taken the gas to be pure ionized hydrogen. Note that
$P_{\rm gas}/P_{\rm rad} = 8/(s_{\rm rad}/k) \ll 1$.
The energy functional of a star now becomes
\begin{equation} \label{en2}
E(\rho_c) = U_{\rm rad} + W + \Delta U_{\rm gas} + \Delta W_{\rm PN},
\end{equation}
where $\Delta U_{\rm gas}$ is the internal energy perturbation due to thermal
gas energy and $\Delta W_{\rm PN}$ is the PN perturbation to the gravitational 
potential energy.  Applying the equilibrium condition~(\ref{1var}) now yields
$M_{\rm eq} \approx M_{\rm eq}^{\rm Newt}$ times a slowly varying function of 
$\rho_c$  (see Fig.~\ref{sketch}{\it b}).  The turning point on the 
equilibrium curve marks the onset of radial instability; the marginally stable 
critical configuration is characterized by
\begin{eqnarray} \label{crit1}
\rho_{c,{\rm crit}} & = & 2 \times 10^{-3} M_6^{-7/2} 
{\rm gm\ cm^{-3}}, \nonumber \\
T_{c,{\rm crit}} & = & (3 \times 10^7) M_6^{-1}~{\rm K}, \\
(R/M)_{\rm crit} & = & 1.6 \times 10^3 M_6^{1/2}, \nonumber 
\end{eqnarray}
where $M_6$ denotes the mass in units of $10^6 ~M_{\odot}$.  (Here and 
throughout we adopt gravitational units and set $G=1=c$.)

Finally, let us consider a uniformly rotating SMS spinning at the 
mass-shedding limit \cite{baum99}.  A centrally condensed object 
like an $n=3$ polytrope can only support a small amount of rotation before 
matter flys off at the equator. At the mass-shedding limit, the ratio of 
rotational kinetic to gravitational potential energy is only 
$T/|W| = 0.899 \times 10^{-2} \ll 1$. Most of the mass resides in a nearly 
spherical interior core, while the low-mass (Roche) envelope bulges
out in the equator: $R_{\rm eq}/R_{\rm pole} = 3/2$. When we include the
contribution of rotational kinetic energy to the energy functional, we must
now also include the effects of relativistic gravity to Post-Post-Newtonian 
(i.e. PPN) order, since both $T$ and $\Delta W_{\rm PN}$ scale with $\rho_c$
to the same power. The energy functional becomes
\begin{equation} \label{en3}
E(\rho_c) = U_{\rm rad} + W + \Delta U_{\rm gas} + \Delta W_{\rm PN}
+ \Delta W_{\rm PPN} + T
\end{equation}
Applying the equilibrium condition~(\ref{1var}), holding $M$, angular momentum 
$J$ and $s$ fixed, now yields $M_{\rm eq} \approx M_{\rm eq}^{\rm Newt}$ times 
a slowly varying function of $\rho_c$ (see Fig.~\ref{sketch}{\it c}).  If we 
restrict our attention to rapidly rotating stars with $M > 10^5 ~M_{\odot}$ 
(the typical size of the seed black holes adopted in some recent galaxy merger 
simulations; see, e..g., \cite{haehnelt00}) the influence of thermal gas pressure is unimportant in determining the 
critical point of instability.  The turning point on the equilibrium curve 
then shifts to higher density and compaction than the critical values for 
nonrotating stars, reflecting the stabilizing role of rotation:
\begin{eqnarray} \label{crit2}
\rho_{c,{\rm crit}} & = & 0.9 \times 10^{-1} M_6^{-2} 
{\rm gm\ cm^{-3}}, \nonumber \\
T_{c,{\rm crit}} & = & (9 \times 10^7) M_6^{-1/2}~{\rm K}, \\
(R_{\rm pole}/M)_{\rm crit} & = & 427, \nonumber \\
(J/M^2)_{\rm crit} & = & 0.97. \nonumber 
\end{eqnarray}
The actual values quoted above for the critical configuration were determined 
by a careful numerical integration of the general relativistic equilibrium 
equations for rotating stars \cite{baum99}; they are in close 
agreement with those determined analytically by the variational treatment. The 
numbers found for the nondimensional critical compaction and angular momentum 
are quite interesting.  First, they are universal ratios that are independent 
of the mass of the SMS. This means that a single relativistic simulation will 
suffice to track the collapse of a marginally unstable, maximally rotating SMS 
of {\it arbitrary} mass.  Second, the large value of the critical radius shows 
that a marginally unstable configuration is nearly Newtonian at the onset of 
collapse. Third, the fact that the angular momentum parameter of the critical 
configuration $J/M^2$ is below unity suggests that, in principle, the entire 
mass and angular momentum of the configuration could collapse to a rotating 
black hole in vacuum without violating the Kerr limit for 
black hole spin (but see the next section below!).

\section*{The Outcome of Collapse}

There are several plausible outcomes that one might envision {\it a priori}\ 
for the dynamical collapse of a uniformly rotating SMS once it reaches the 
marginally unstable critical point identified above. It could collapse to a 
clumpy, nearly axisymmetric disk, similar to the one arising in the Newtonian 
SPH simulations for the isothermal ($\Gamma=1$) 
implosion of an initially homogeneous, uniformly rotating, low-entropy cloud 
\cite{loeb94}. 
Alternatively, the disk might develop a large-scale,  nonaxisymmetric bar. 
After all, the onset of a dynamically unstable bar mode in a spinning 
equilibrium star occurs when the ratio $T/|W| \approx 0.27$ (see, e.g., 
\cite{chandra69} and \cite{lai93} for Newtonian treatments
and \cite{saijo01} and \cite{shibata00a} for simulations 
in general relativity). Since $T/|W|$ is $0.899 \times 10^{-2}$ at the onset 
of collapse and scales roughly as $R^{-1}$ during collapse, assuming 
conservation of mass and angular momentum, this ratio climbs 
above the dynamical bar 
instability threshold when the SMS collapses to $R/M \approx 20$, well before 
the horizon is reached.  The growth of a bar might begin at this point.  
Indeed, a weak bar forms in simulations of rotating supernova core collapse 
\cite{rampp98, brown01}, but in supernova the equation of state 
stiffens ($\Gamma > 4/3$) at the end of the collapse, triggering a bounce and 
thereby allowing more time for the bar to develop.  A rapidly rotating 
unstable SMS might not form a disk at all, but instead collapse entirely to a 
Kerr black hole; not surprisingly, a nonrotating spherical SMS has been shown 
to collapse to a Schwarzschild black hole \cite{shap79}. 
Alternatively, the unstable rotating SMS might collapse to a rotating black 
hole {\it and}\ an ambient disk.

\begin{figure}[tb]
\begin{tabular}{rl}
\hspace*{2.5cm}\includegraphics[width=4.5cm]{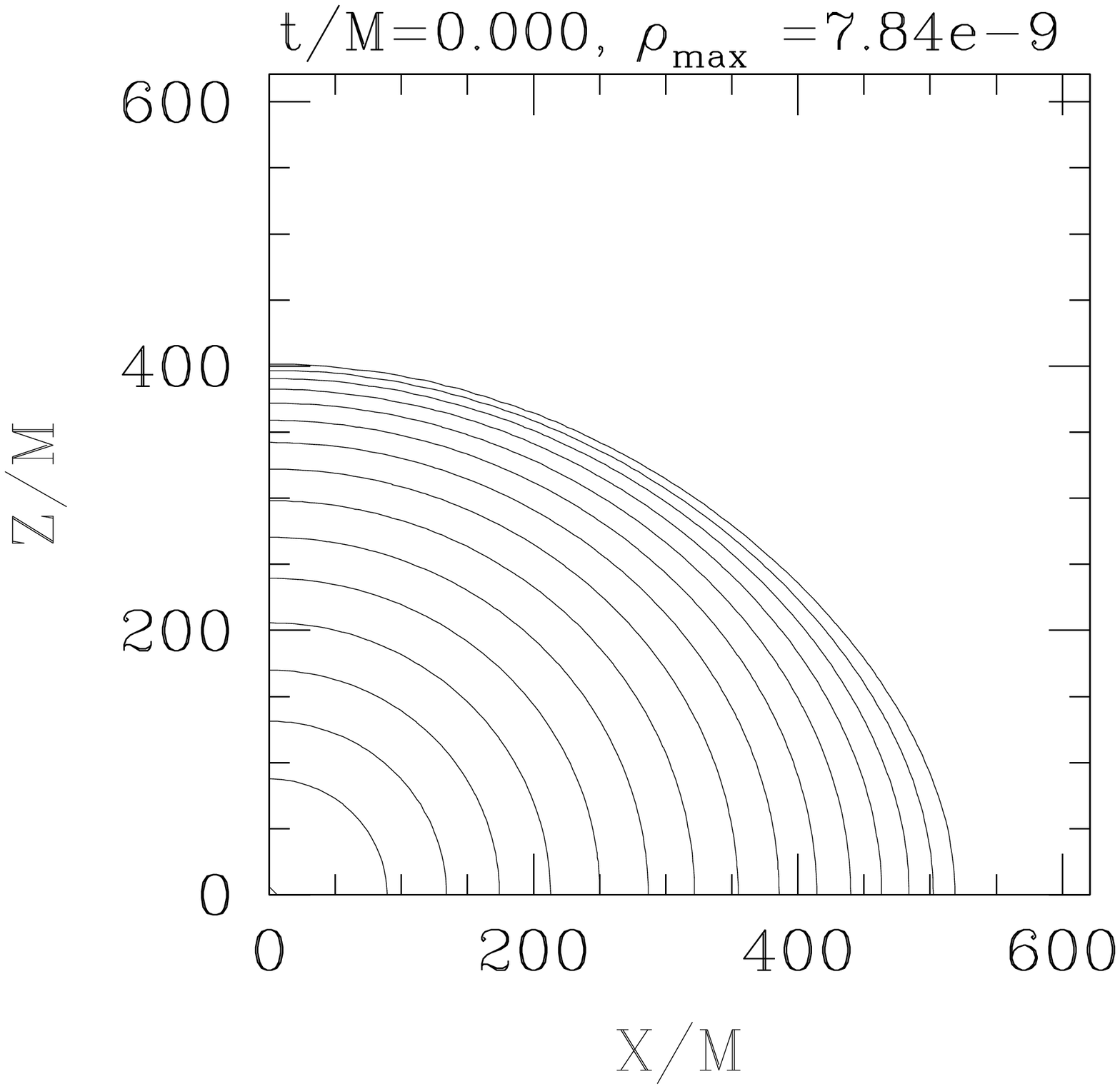}
&\includegraphics[width=4.5cm]{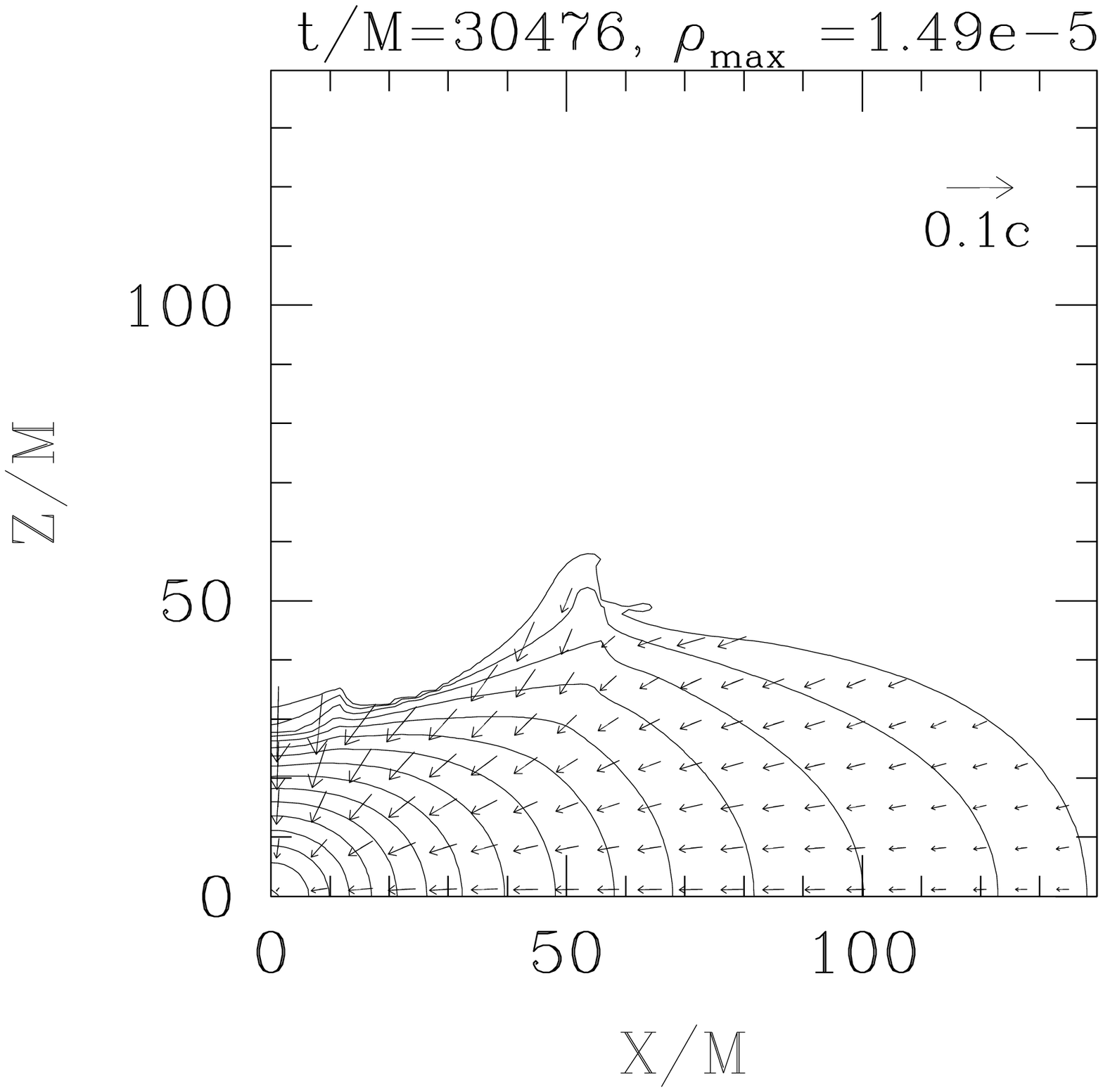}\\
\hspace*{2.5cm}\includegraphics[width=4.5cm]{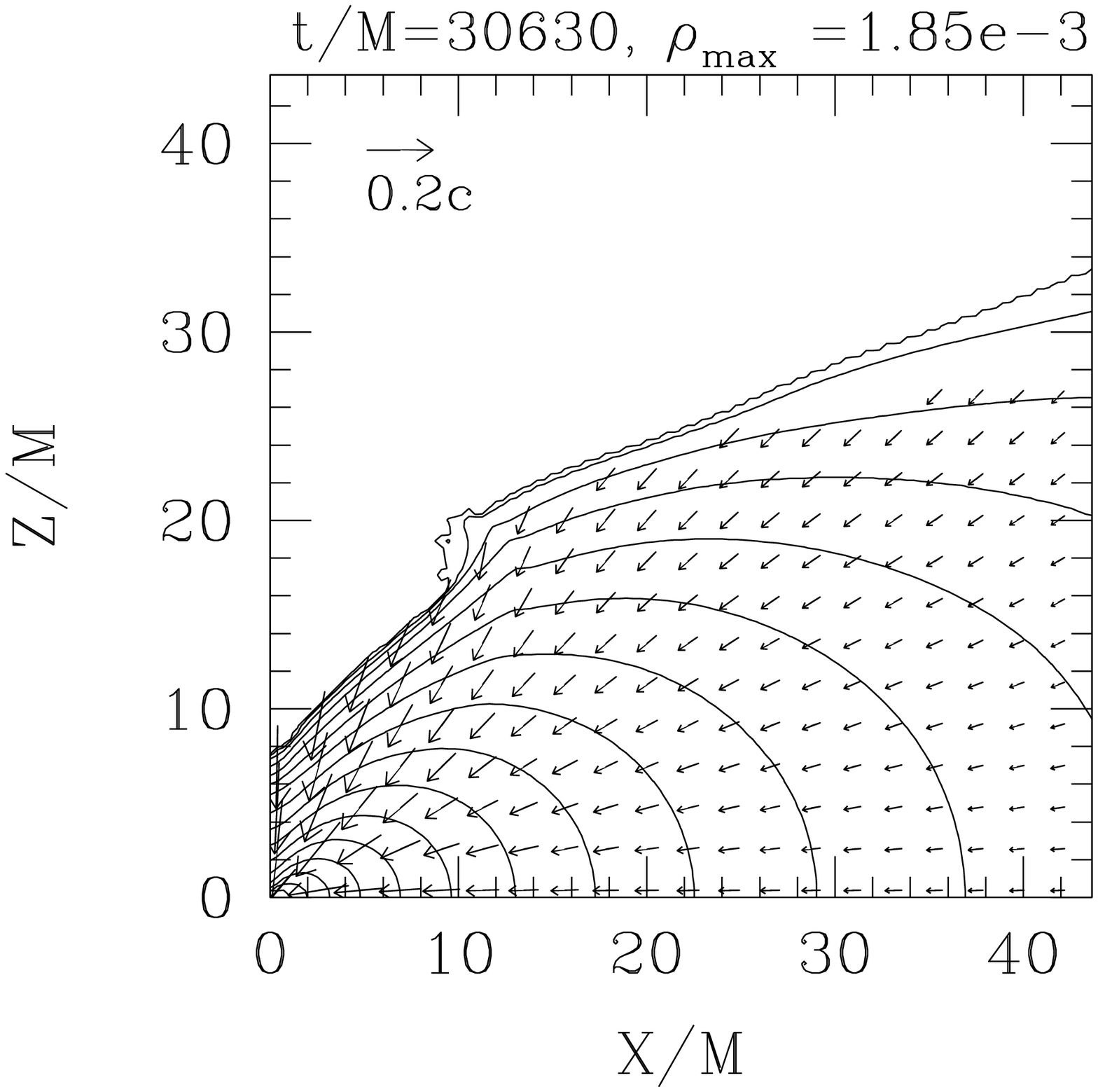}
&\includegraphics[width=4.5cm]{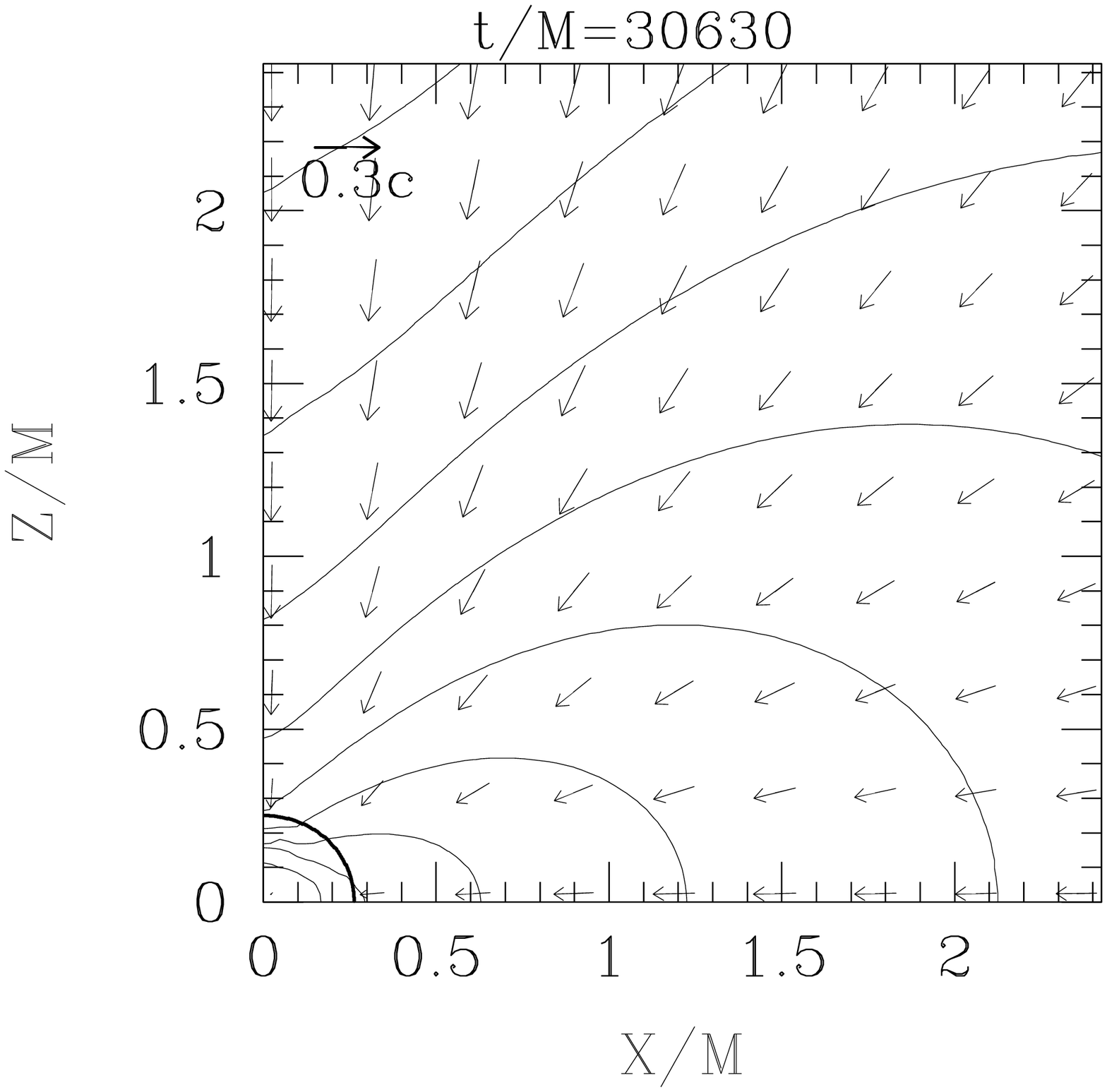}
\end{tabular}
\caption{Snapshots of density and velocity profiles during the implosion
of a marginally unstable SMS of arbitrary mass $M$ rotating uniformly at 
break-up speed at $t=0$.  The contours are drawn for $\rho/\rho_{\rm max}
=10^{-0.4j}~(j=0-15)$, where $\rho_{\rm max}$ denotes the maximum density at 
each time.  The fourth figure is the magnification of the third one in the 
central region: the thick solid curve at $r \approx 0.3M$ denotes the location 
of the apparent horizon of the emerging SMBH. 
(From Shibata \& Shapiro \cite{shibata02})}
\label{SMS}
\end{figure}

\begin{figure}[tb]
\centering
\includegraphics[width=6cm]{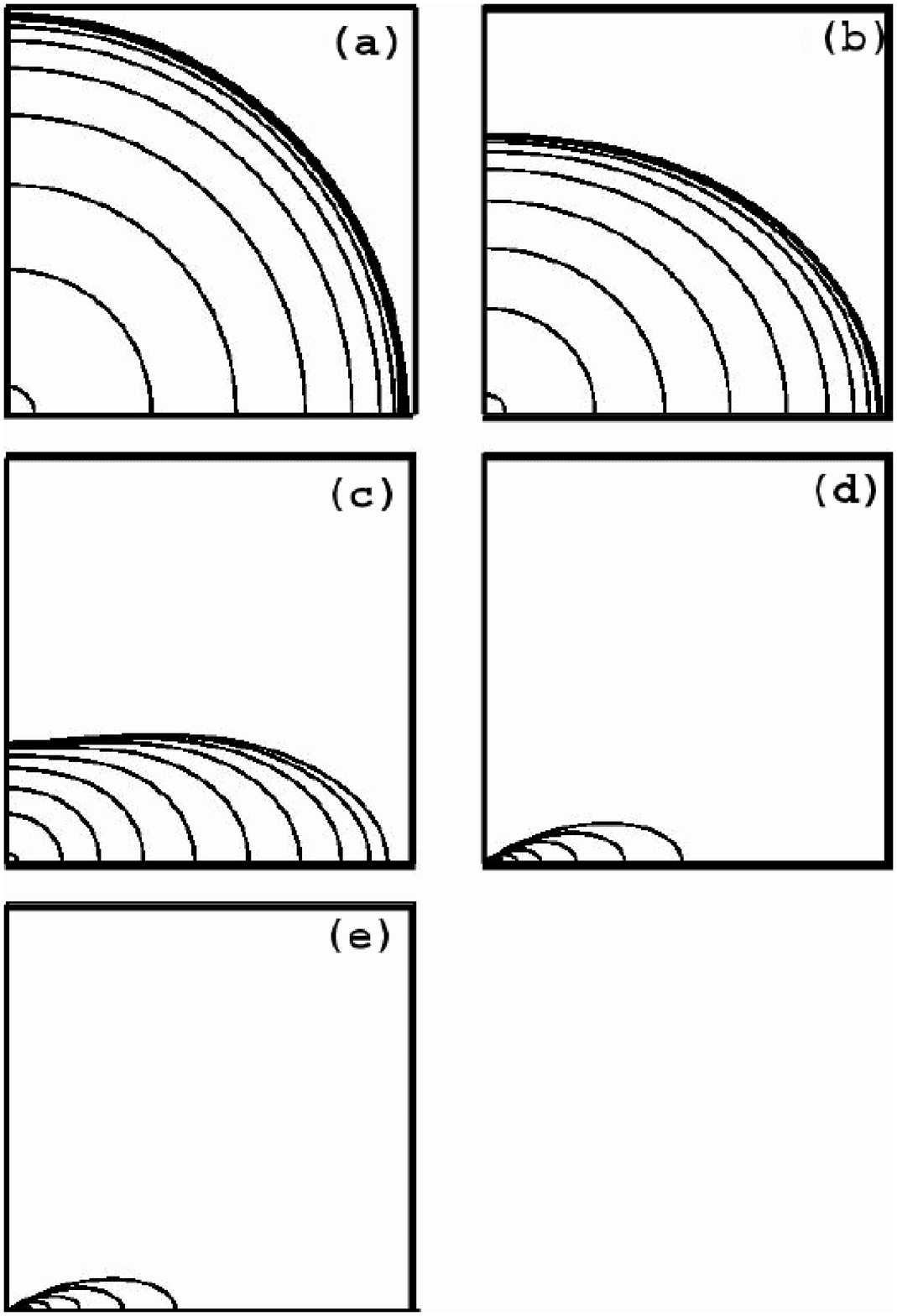}
\caption
{Quasi-stationary evolution snapshots of a SMS that cools and contracts
in the absence of any viscosity or magnetic fields 
from a nearly spherical initial state that is in very slow, uniform rotation.
The maximum density is normalized to unity; the highest density contour
level is 0.9 and subsequent levels range from $10^{-1}$ to $10^{-10}$ and
are separated by a decade. The final configuration has reached the mass-shedding
limit and is toroidal and differentially rotating, with $T/|W| \approx 0.27$ and 
$\Omega_{\rm pole}/ \Omega_{\rm eq} \approx 2.5 \times 10^3$.
(From New \& Shapiro \cite{new01}}
\label{diff}
\end{figure}

Recently we have performed two simulations that together resolve 
the fate of a marginally unstable, 
maximally rotating SMS of arbitrary mass $M$. In \cite{saijo02}
we followed the collapse in full $3D$, but assumed PN theory. 
We tracked the implosion up to 
the point at which the central spacetime metric begins to deviate appreciably
from flat space at the stellar center. We found that the massive core 
collapses homologously during the Newtonian epoch of collapse, and that 
axisymmetry is preserved up to the termination of the integrations.  This 
calculation motivated us \cite{shibata02} to follow the collapse in full 
general relativity by assuming axisymmetry from the beginning to maximize 
spatial resolution (see Fig.~\ref{SMS}).  We found that the 
final object is a Kerr-like black hole 
surrounded by a disk of orbiting gaseous debris.  The final black hole mass 
and spin were determined to be  $M_h/M \approx 0.9$ and $J_h/M_h^2 \approx 
0.75$. The remaining mass goes into the disk of mass $M_{\rm disk}/M \approx 
0.1$.  A disk forms even though the total spin of the progenitor star is 
safely below the Kerr limit.  This outcome results from the fact that the dense 
inner core collapses homologously to form a central black hole, while the 
diffuse outer envelope avoids capture because of its high angular momentum. 
Specifically, in the outermost shells, the angular momentum per unit mass $j$, 
which is strictly conserved on cylinders, exceeds $j_{\rm ISCO}$, the specific 
angular momentum at the innermost stable circular orbit about the final hole. 
This fact suggests how the final black hole and disk parameters can be 
calculated {\it analytically}\ from the initial SMS density and angular 
momentum distribution \cite{shap02}. The result applies to the 
collapse of {\it any}\ marginally unstable $n=3$ polytrope at mass-shedding. 
Maximally rotating stars which are characterized by stiffer equations of state 
and smaller $n$ (higher $\Gamma$) do not form disks, typically, since they are
more compact and less centrally condensed at the onset of collapse \cite{shibata00b}.

\begin{figure}[t]
\centering
\includegraphics[width=9cm]{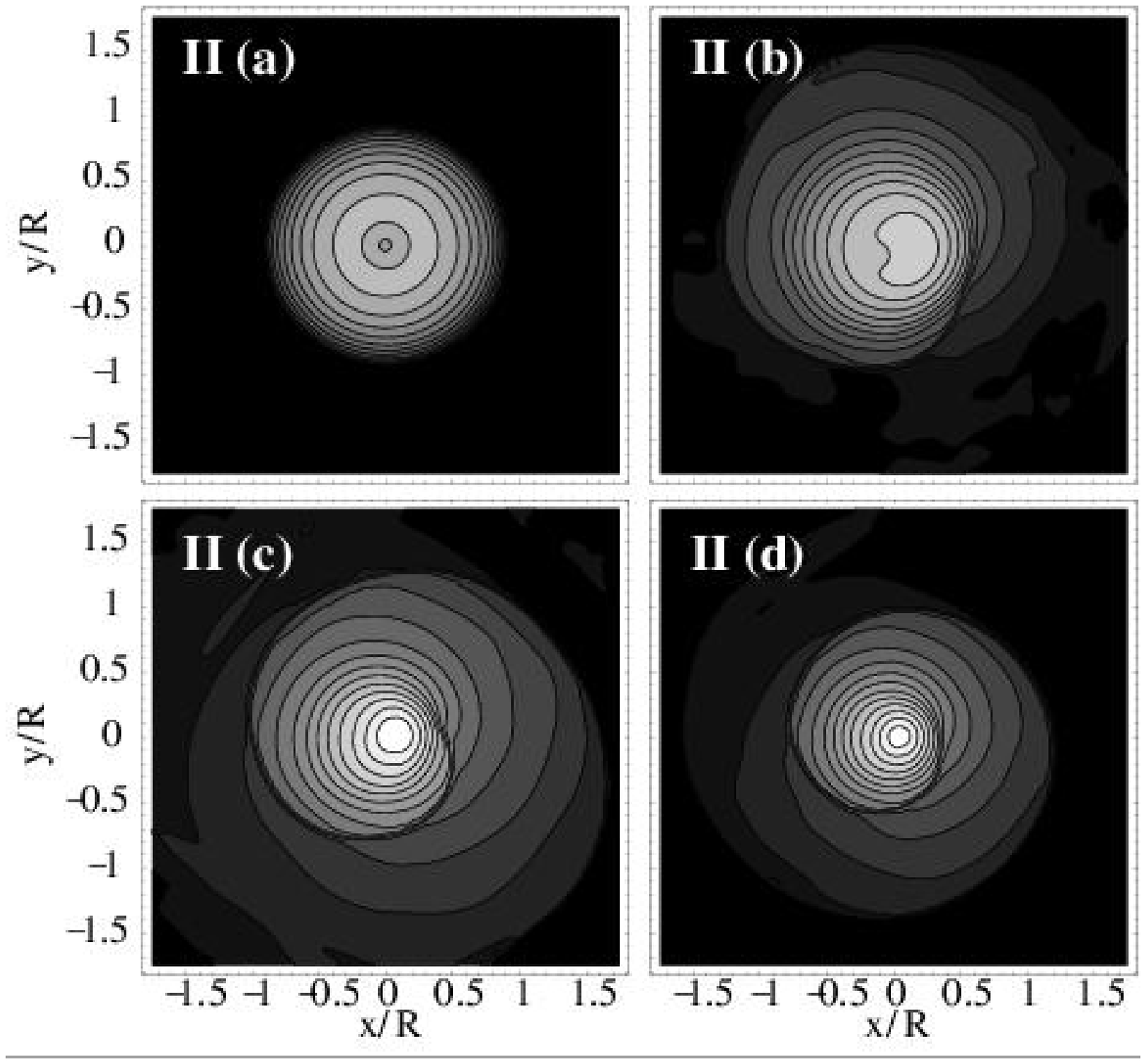}
\caption
{Snapshots of density contours in the equatorial plan of differentially
rotating Newtonian toroids of different polytropic indicies, n. All of the stars
are constructed from the same differential rotation law with
$\Omega_{\rm pole}/\Omega_{\rm eq} \approx 26$ and $T/|W| \approx 0.14$
at $t=0$.  The snapshots are taken after the stars have evolved for
many central rotation periods, $P_c$.
Stars with higher n exhibit a one-armed spiral m=1 instability.
The panels show results for $(n, t/P_c) = (a) (2,37); (b) (2.5,24);
(c) (3,17)$; and $(d) (3.33, 19)$.
The contour lines denote densities
$\rho/\rho_{\rm c} = 10^{- (16-i) d}  (i=1, \cdots, 15)$.
(From Saijo, Baumgarte \& Shapiro \cite{saijo03})}
\label{spiral}
\end{figure}

\begin{figure}[t]
\centering
\includegraphics[width=9cm]{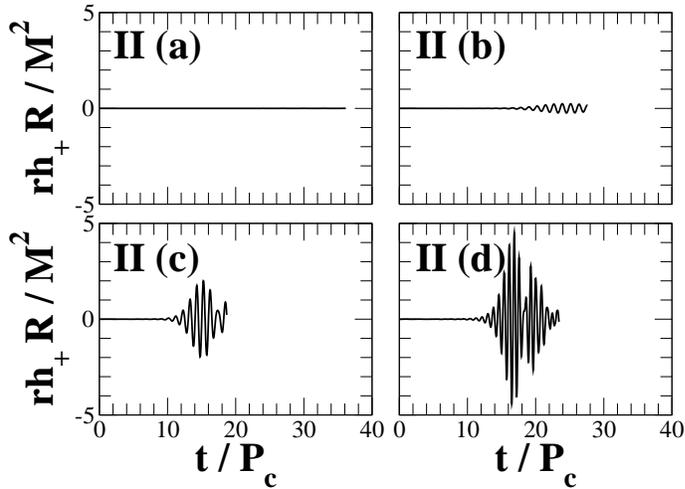}
\caption
{Gravitational radiation waveforms as seen by a distant observer on the
rotation axis for the configurations shown in Fig.~\ref{spiral}. The
radius $r$ is the distance to the source and the ratio $M/R$ is the
(arbitrary) compaction of the initial configuration.
(From Saijo, Baumgarte \& Shapiro \cite{saijo03})}
\label{wave}
\end{figure}

The above calculations show that a SMBH formed from the collapse of a 
maximally rotating SMS is always born with a ``ready-made'' accretion disk. 
This disk might provide a convenient source of fuel to power the central 
engine.  The calculations also show that the SMBH will be born rapidly 
rotating.  This fact is intriguing in light of suggestions that observed SMBHs 
rotate rapidly (e.g., \cite{wilms01,elvis02}).

The implosion will result in a {\it burst} of gravitational
waves. Current relativistic collapse calculations for this problem 
break down soon after the apparent horizon of the hole approaches its asymptotic size.
Moreover, the need to resolve the centrally condensed inner core precludes 
extending the numerical grid very far out into the wave zone. These two 
restrictions prevent a
reliable determination of the complete burst waveform at present. Crude estimates
can be generated for the frequency and amplitude 
based on the quadrupole formula, and the results suggest that 
they are well within the range of detectability for LISA for a 
reasonable spectrum of masses. More interesting, perhaps, is 
the possible generation of {\it quasi-periodic}
gravitational waves from nonaxisymmetric instabilities that might 
arise in the ambient disk; the characteristics of these waves are
likely to be comparable those discussed in the next section and in 
Eqn.~(\ref{hplus}) below.

\section*{An Alternative Scenario}

The previous analysis assumes that a combination of turbulent viscosity and
magnetic fields combine to keep a spinning SMS rotating {\it uniformly} prior to
the onset of collapse. While this is the most likely possibility 
\cite{bisno67, wag69, zel71, shap00}, New and Shapiro \cite{new01} have discussed
an alternative. In the absence of viscosity or magnetic fields, 
an evolving SMS will conserve angular momentum on cylinders as it cools and 
contracts in a quasi-stationary manner. So even if it is rotating uniformly at formation,
it will evolve to a state of {\it differential} rotation. Our 
evolution calculations
indicate that an axisymmetric configuration that is slowly 
rotating uniformly and nearly spherical
at birth ($T/|W| \ll 1$) will evolve to a differentially rotating toroid 
(see Fig.~\ref{diff}).
Prior to encountering any relativistic radial instability, 
the star reaches the mass-shedding limit, at which point $T/|W| \approx 0.27$ and  
differential rotation is
extreme, $ \Omega_{\rm pole}/ \Omega_{\rm eq} \approx 2.5 \times 10^3$.

Toroidal configurations which have high values of $T/|W|$ and substantial 
differential rotation and central mass concentration are subject to nonaxisymmetric
dynamical instabilities, especially the one-armed spiral $m=1$ instability
\cite{cent01, saijo03}. Softer equations of state with higher polytropic indicies
are very susceptible to this instability (see Fig.~\ref{spiral}), so this mode is
likely to be triggered in a rotating SMS should it evolve with low viscosity and
magnetic field. Such a mode typically induces a secondary 
m=2 bar mode of smaller amplitude, and the bar mode can excite quasi-periodic 
gravitational waves (see Fig.~\ref{wave}). Typical wave amplitudes and wave 
frequencies for a source
at a distance $r$ are given by  
\begin{eqnarray} \label{hplus}
h & \sim & 10^{-16} (T/|W|)(M/R)M_6 r_{\rm GPC}^{-1} 
\sim 10^{-20}M_6r_{\rm GPC}^{-1} , \\
f_{\rm GW}  & \sim & 10^{-1} (T/|W|)^{1/2}(M/R)^{3/2}M_6^{-1} Hz 
\sim 10^{-4} M_6^{-1} Hz. \nonumber 
\end{eqnarray}
where we have set $T/|W| \sim 0.27$ and $R/M \sim 10^3$ in evaluating the
above quantities. These values also apply to waves generated 
by bar modes that may arise in a similar fashion in the
ambient disk surrounding the SMS black hole that forms from the collapse of
a relativistically unstable, uniformly rotating SMS (see discussion above).
Both the amplitudes and frequencies put these waves well within the detectable 
range for LISA for a realistic distribution of masses. Future simulations 
that we are planning will
explore this possibility in greater detail, and help resolve which scenario
best describes the final fate of a rotating SMS.

\section*{acknowledgments}
This work was supported in part by NSF Grants PHY-0090310 and PHY-0205155 and
NASA Grant NAG5-10781 at the University of Illinois at
Urbana-Champaign.




\begin{thebibliography}{999}

\bibitem{abel00}
Abel, T., Bryan, G. L., \& Norman, M. L. 2000, \apj, 540, 39

\bibitem{baum99}
Baumgarte, T. W., \& Shapiro, S. L. 1999, \apj, 526, 941

\bibitem{baum03}
Baumgarte, T. W., \& Shapiro, S. L. 2003, Phys. Reports, 376/2, 41

\bibitem{bisno67}
Bisnovatyi-Kogan, G. S., Zel'dovich, Ya. B., \& Novikov, I. D. 1967, 
Soviet Astron., 11, 419

\bibitem{brom99}
Bromm, V., Coppi, P. S., \& Larson, R. B. 1999, \apj, 527, L5

\bibitem{brom03}
Bromm, V., \& Loeb, A. 2003, \apj, in press (astro-ph/0212400)

\bibitem{brown01}
Brown, J. D.  2001, in Astrophysical Sources for Ground-based Gravitational 
Wave Detectors, ed. J. M. Centrella (New York: AIP), 234

\bibitem{cent01}
Centrella, J. M., New, K. C. B., Lowe, L. L. \& Brown, J. D. 2001,
\apj, 550, L193 

\bibitem{chandra64a}
Chandrasekhar, S. 1964a, Phys. Rev. Lett., 12, 114, 437E

\bibitem{chandra64b}
Chandrasekhar, S. 1964b, \apj, 140, 417

\bibitem{chandra69}
Chandrasekhar, S.  1969, Ellipsoidal Figures of Equilibrium 
(New Haven: Yale Univ. Press)

\bibitem{elvis02}
Elvis, M., Risaliti, G., \& Zamorani, C. 2002, \apj, 565, L75

\bibitem{fan00}
Fan, X., et al. 2000, \aj, 120, 1167

\bibitem{fan01}
Fan, X., et al. 2001, \aj, 122, 2833

\bibitem{feyn}
Feynmann, R. unpublished, as quoted in \cite{fowl64}

\bibitem{fowl64}
Fowler, W. A. 1964, Rev. Mod. Phys., 36, 545, 1104E

\bibitem{fry01}
Fryer, C. L., Woosley, S. E., \& Heger, A. 2001, \apj, 550, 372

\bibitem{full86}
Fuller, G. M., Woosley, S. E., \& Weaver, T. A. 1986, \apj, 307

\bibitem{genz97}
Genzel, R., Eckart, A., Ott, T., \& Eisenhauer, F. 1997, \mnras, 291, 219

\bibitem{ghez00}
Ghez, A. M., Morris, M., Becklin, E. E., Tanner, A., \& Kremenek, T. 2000, 
Nature, 407, 349
 
\bibitem{gned01}
Gnedin, O. Y. 2001, Class. \& Quant. Grav., 18, 3983

\bibitem{haehnelt00}
Haehnelt, M. G., \& Kauffmann, G. 2000, \mnras, 318, L35

\bibitem{ho99}
Ho, L.~C. 1999, in Observational Evidence for Black Holes in the Universe,
ed. S.~K. Chakrabarti (Dordrecht: Kluwer), 157

\bibitem{lai93}
Lai, D., Rasio, F. A, \& Shapiro, S. L. 1993, \apjs, 88, 205

\bibitem{loeb94}
Loeb, A., \& Rasio, F. A. 1994, \apj, 432, 52

\bibitem{mach99}
Macchetto, F. D. 1999, in Towards a New Millennium in Galaxy Morphology,
ed. D. L. Block et al. (Dordrecht: Kluwer)

\bibitem{new01}
New, K. C. B., \& Shapiro, S. L. 2001, \apj, 548, 439

\bibitem{rampp98}
Rampp, M., M\"uller, E., \& Ruffert, M. 1998, \aa, 332, 969

\bibitem{rasio99}
Rasio, F. A., \& Shapiro, S. L. 1999, Class. Quant. Grav., 16, 1

\bibitem{rees84}
Rees, M. J. 1984, \annrev, 22, 471

\bibitem{rees98}
Rees, M. J. 1998, in Black Holes and Relativistic Stars, ed.  R. M. Wald (Chicago: 
Chicago Univ. Press), 79

\bibitem{rees01}
Rees, M. J. 2001, in Black Holes in Binaries and Galactic Nuclei, ed.  L. Kaper, 
E. P. J. van den Heurel, \& P. A. Woudt (New York: Springer-Verlag), 351

\bibitem{rich98}
Richstone, D., et al. 1998, Science, 395, A14

\bibitem{saijo03}
Saijo, M., Baumgarte, T. W., \& Shapiro, S. L. 2003, submitted to \apj, 
(astro-ph/0302436)

\bibitem{saijo01}
Saijo, M., Shibata, M., Baumgarte, T. W., \& Shapiro, S. L. 2001, \apj, 548, 919

\bibitem{saijo02}
Saijo, M., Shibata, M., Baumgarte, T. W., \& Shapiro, S. L. 2002, \apj, 569, 349

\bibitem{scho02}
Sch\"odel, R., et al. 2002, Nature, 419, 694

\bibitem{shap03}
Shapiro, S. L. 2003, in Carnegie Observatories Astrophysics Series, Vol 1:
Coevolution of Black Holes and Galaxies, ed.  
L. C. Ho (Cambridge: Cambridge Univ. Press), in press (astro-ph/0304202).

\bibitem{shap00}
Shapiro, S. L. 2000, \apj, 544, 397

\bibitem{shap02}
Shapiro, S. L., \& Shibata, M. 2002, \apj, 577, 904

\bibitem{shap79}
Shapiro, S. L., \& Teukolsky, S. A. 1979, \apj, 234, L177

\bibitem{shap83}
Shapiro, S. L., \& Teukolsky, S. A.  1983, Black  Holes, White Dwarfs, 
and Neutron Stars: The Physics of Compact Objects (New York: Wiley Interscience)

\bibitem{shibata00a}
Shibata, M., Baumgarte, T. W., \& Shapiro, S. L. 2000a, \apj, 542, 453

\bibitem{shibata00b}
Shibata, M., Baumgarte, T. W., \& Shapiro, S. L. 2000b, \prd, 61, 44012

\bibitem{shibata02}
Shibata, M., \& Shapiro, S. L. 2002, \apj, 527, L39

\bibitem{sol82}
Soltan, A. 1982, \mnras, 200, 115

\bibitem{wag69}
Wagoner, R. V. 1969, \annrev, 7, 553

\bibitem{wilms01}
Wilms, J., Reynolds, C. S., Begelman, M. C., Reeves, J., Molendi, S., 
Staubert, R., \& Kendziorra, E. 2001, \mnras, 328, L27

\bibitem{yu02}
Yu, Q., \& Tremaine, S., 2002, \mnras, 335, 965

\bibitem{zel71}
Zel'dovich, Ya. B., \& Novikov, I. D. 1971, Relativistic Astrophysics, Vol.~1 
(Chicago: Univ. of Chicago Press)

\end{thebibliography}
\end{document}